\documentclass[a4paper,11pt,aps,pra,amsmath,amssymb,showpacs]{revtex4}

\usepackage{latexsym}
\usepackage[latin1]{inputenc}
\usepackage{bbm}

\newcommand{\up}[1][]{_{\uparrow #1}}
\newcommand{\down}[1][]{_{\downarrow #1}}

\renewcommand{\vec}[1]{\mathbf{#1}}

\usepackage[dvips]{graphicx}

\begin{document}

\title{Sound velocity and dimensional crossover in a superfluid Fermi
gas in an optical lattice}
\date{\today}
\author{T. Koponen$^1$}
\author{J.-P. Martikainen$^{1-2}$}
\author{J. Kinnunen$^1$}
\author{P. T\"{o}rm\"{a}$^1$}
\affiliation{$^1$ Nanoscience Center, Department of Physics, P.O. Box 35, FI-40014 University
  of Jyv\"{a}skyl\"{a}, Finland \\$^2$ Department of Physical Sciences, 
P.O. Box 64, FI-00014 University of Helsinki,  Finland}
\begin{abstract}
We study the sound velocity in cubic and non-cubic three-dimensional optical
lattices. We show how the van Hove singularity of the free Fermi gas
is smoothened by interactions and eventually vanishes when interactions
are strong enough. For non-cubic lattices, we show that the
speed of sound (Bogoliubov-Anderson phonon) shows clear signatures of dimensional crossover
both in the 1D and 2D limits. 
\end{abstract}
\maketitle

\section{Introduction}
\label{sec:intro}
The experimental ability to tune and control the quantum properties of
matter
has increased dramatically during the past decade.
The evaporative cooling of bosons enabled the creation
of Bose-Einstein condensates in dilute atomic
gases~\cite{Anderson1995a,Davis1995a,Bradley1995a}.
Due to statistics fermionic atoms proved harder to cool, but the
favorable collisional
properties of certain two-component mixtures made the efficient cooling
of
fermions into quantum degeneracy
possible~\cite{DeMarco1999a,Truscott2001a,Granade2002a}. 

If two different kinds of fermions
are mixed, they can interact via s-wave interactions. If this
interaction is attractive, fermions can pair and form a superfluid
as predicted by the Bardeen-Cooper-Schrieffer (BCS)
theory~\cite{Stoof1996b}.
However, the transition temperature of the superfluid transition is
typically much smaller than degeneracy temperature of the
fermions and therefore hard to reach experimentally. This difficulty 
was recently overcome by using the so-called Feshbach resonance
to change the strength of the interaction between two atoms.

Near a Feshbach resonance an external magnetic field~\cite{Duine2004a}
is used to tune the energy difference between the molecular bound state
and the two-atom continuum. Changes in this energy difference 
give rise to a resonant behavior of the interaction strength between two
atoms. In particular, 
the interaction strength can be varied from a small and
negative value (BCS-side),  into a very large value
close to resonance,
and finally into a small and positive value (BEC-side) at the other side
of the resonance. In the Bose-Einstein condensate side of the resonance 
the system forms diatomic 
molecules~\cite{Regal2003a,Strecker2003a,Cubizolles2003a}
and, at sufficiently low temperature, a condensate of
molecules~\cite{Jochim2003b,Greiner2003a,Zwierlein2003a,Bourdel2004a}. 
Fermion pairing 
is possible on the BCS-side of the resonance. Strong indications of such
pairing
were experimentally observed employing magnetic field sweeps across 
the resonance~\cite{Regal2004a,Zwierlein2004a} and 
by monitoring the behavior of collective
modes~\cite{Kinast2004a,Bartenstein2004b}.
Soon after the pairing gap was directly measured using  radio-frequency
spectroscopy~\cite{Chin2004a,Kinnunen2004b}. Finally, the smoking gun
of superfluidity, a vortex lattice, in a fermionic system was very
recently observed~\cite{Zwierlein2005a}.

New tools to manipulate the atomic clouds have been developed.
In particular, optical lattices
have proved to be  an especially useful tool. In the context
of degenerate quantum gases,
optical lattices~\cite{Orzel2001a,Greiner2001b,Burger2001a,Hadzibabic2004a} have
been used to explore, for example, 
non-classical states, phase coherence, condensate 
dynamics, as well as matter wave interference.
Since both the lattice depth
and the lattice geometry can be easily changed, there is 
the possibility of investigating experimentally anisotropic quantum
systems as well as quantum phase transitions from, for example, a
superfluid into an insulator as the tunneling strength between nearest
neighbors is reduced~\cite{Greiner2002a}.
Also, sufficiently deep lattices in certain directions can be used to
control the dimensionality of the system.

The ability to trap atoms into optical lattices does not
depend on which statistics, fermionic or bosonic, the atoms
obey. Indeed, recently degenerate fermions have been studied 
experimentally in one-dimensional~\cite{Modugno2003a,Pezze2004a} as well as in
three-dimensional~\cite{Kohl2005a,Stoferle2005a} optical 
lattices.
There has also been rapid progress on the theoretical front.
Among other things, there are several studies on fermions and
boson-fermion mixtures in optical
lattices~\cite{Hofstetter2002a,Roth2003b,Carr2005b,Lewenstein2004a,Wang2004a,Dickerscheid2005b}
and
on fermion dynamics in optical
lattices~\cite{Wouters2004a,Rodriguez2004a,Pitaevskii2005a}.
In addition, more exotic fermionic systems with several 
flavors have been discussed~\cite{Honerkamp2004a}.

In this paper we calculate the density response of superfluid
fermions in a three-dimensional (3D) optical lattice,
to elucidate how to investigate dimensional 
crossover effects. Therefore we calculate the above signatures 
also for non-cubic optical lattices with varying lattice depths in
different directions. We show that
the speed of sound (Bogoliubov-Anderson phonon) as a function of filling fraction
behaves qualitatively differently when the dimensionality
of the lattice is changed.

 While we work consistently
within the framework of the BCS theory, we expect our results
and discussions to be instructive also somewhat closer to the Feshbach
resonance where interactions are stronger. Naturally it is to be
understood, that if the scattering length
characterizing the interaction strength becomes comparable
to the lattice spacing, the continuum Feshbach physics will 
change~\cite{Wouters2003a,Dickerscheid2005a,Orso2005a}. If we furthermore have
about one atom per lattice site, the simple BCS description is expected
to fail in this case. 

This paper is organized as follows. In sec.~\ref{sec:BCS} we
review the standard BCS-theory. In section \ref{sec:DR} we discuss
the density response and how it is calculated. In section
\ref{sec:SoS} we present and analyze our results on the speed of first
sound in cubic lattices and in section \ref{sec:AIS} study
the dimensional crossover in non-cubic lattices. 
Conclusions are presented in sec.~\ref{sec:conclusions}. 

\section{BCS-theory}
\label{sec:BCS}
The single-band Hubbard Hamiltonian in the context of ultracold gases can be derived from the full microscopic
Hamiltonian~\cite{Jaksch1998a}. To achieve this, one must expand the
field operators of the microscopic Hamiltonian in the basis of the
well localized Wannier functions, assume that the optical lattice is
sufficiently deep, and that the gas is sufficiently dilute. Then
it is possible neglect all but the nearest neighbour interactions.
Furthermore, assuming that the temperature and interactions are low
enough so that there are no excitations to the excited oscillator
levels of the individual lattice wells, the Hamiltonian becomes, for fermions, the single-band Fermi-Hubbard Hamiltonian:
\begin{equation}
\begin{split}
H -\mu N=& -\mu\sum_n \left(\hat{c}\up[n]^\dagger \hat{c}\up[n] +
\hat{c}\down[n]^\dagger \hat{c}\down[n] \right)+ U\sum_n
\hat{c}\up[n]^\dagger \hat{c}\down[n]^\dagger \hat{c}\down[n]\hat{c}\up[n]\\
& -\left(J_x \sum_{\langle n,m\rangle_x}  + J_y \sum_{\langle
  n,m\rangle_y} + J_z \sum_{\langle
  n,m\rangle_z}\right)
\left(\hat{c}\up[m]^\dagger \hat{c}\up[n] + \hat{c}\down[m]^\dagger \hat{c}\down[n] \right),
\end{split}
\end{equation}
Here the operator \(\hat{c}_{\sigma n}^\dagger\) creates an atom with
spin \(\sigma\) at the
lattice site \(n\) and \(c_{\sigma n}\) annihilates it. The bracket \(\langle n,m \rangle_{\alpha}\) denotes all the nearest neighbour pairs
in the \(\alpha\)-direction, \(\mu\) is the chemical potential, \(U\) is
the interaction strength of the atoms and \(J_x\), \(J_y\), and
\(J_z\) are tunneling strengths in their respective directions. We
assume that there is an equal amount of spin up and spin down atoms,
so the chemical potentials for both states are equal. In order to allow the lattice to be
non-cubic, separate values for the three tunneling strengths are needed.

The parameters \(J\) and \(U\) can, in
principle, be computed numerically using the lowest
Wannier function, but if one approximates the lattice well by a
harmonic potential they can be determined analytically. The interaction parameter is given by
\begin{equation}
\label{eq:interaction}
U= - \frac{8}{\sqrt{\pi}}\frac{a}{\lambda}\left(\frac{2s_{x}s_{y}s_{z}E_{R}^3\hbar^2}{m\lambda^2}\right)^{\frac{1}{4}},
\end{equation}
while the hopping parameter is given by~\cite{Tsuchiya2005a}
\begin{equation}
J_i = E_{R} e^{-\frac{\pi^2 \sqrt{s_i}}{4}}\left(\left(\frac{\pi^2 s_i}{4}-\frac{\sqrt{s_i}}{2}
  \right)-\frac{1}{2}s_i\left(1 + e^{-\sqrt{s_i}}\right)
\right).
\end{equation}
Here \(\lambda\) is the wavelength of the lattice, \(a\) is the
scattering length, \(E_{R} =
\hbar^2(2\pi/\lambda)^2/2m\) is the recoil energy of the lattice and
\(s_x\), \(s_y\) and \(s_z\) are the lattice heights, in recoil
energies, in different directions. 

The energy separation of the states of individual lattice wells is
\begin{equation}
\hbar \omega = \frac{\hbar^2}{m}\sqrt{\frac{s}{2}}\left(\frac{2\pi}{\lambda}\right)^2,
\end{equation}
where \(s\) is the lattice height. From this and formula Eq. \eqref{eq:interaction} it is possible to
estimate the limits of validity for the single band Hubbard model,
i.e. when thermal excitations or interactions are strong enough for
the second band to be populated. For $^6$Li with \(\lambda = 1030\)~nm
and \(s = 2.5\) these limits are approximately
\(3\ \mu K\) for the temperature and \(-8600\ a_0\) for
the scattering length. This upper limit for the scattering length is
so high that typically the on-site interaction approximation fails before
reaching this limit.

The Hamiltonian in momentum space is
\begin{equation}
\label{eq:k-space hamiltonian}
H = \sum_{\vec{k}}\left(\epsilon_{\vec{k}} - \mu \right)\left( \hat{c}\up[\vec{k}]^\dagger \hat{c}\up[\vec{k}] +
 \hat{c}\down[\vec{k}]^\dagger \hat{c}\down[\vec{k}]\right) + \frac{U}{M}\sum_{\vec{k},\vec{q},\vec{k}'}
\hat{c}\up[\vec{k}'+\vec{k}+\vec{q}]^\dagger
\hat{c}\down[\vec{k}'-\vec{k}-\vec{q}]^\dagger \hat{c}\down[\vec{k}' + \vec{k}] \hat{c}\up[\vec{k}'-\vec{k}],
\end{equation}
where \(M\) is the number of lattice sites and the lattice dispersion
is given by
\begin{equation}
\epsilon_{\vec{k}} = 2J_x(1-\cos(k_x d))+2J_y(1-\cos(k_y d)) + 2J_z(1-\cos(k_z d)).
\end{equation}
Note that the dispersion is even, so \(\epsilon_{\vec{k}}=\epsilon_{-\vec{k}}\).
The Heisenberg equations of motion for the operators \(\hat{c}\up[-\vec{k}]\) and
\(\hat{c}\down[\vec{k}]^\dagger\) are
\begin{equation}
\begin{split}
-i\hbar \frac{\partial}{\partial t} \hat{c}\up[-\vec{k}] &=
[H,\hat{c}\up[-\vec{k}]]\\
-i\hbar \frac{\partial}{\partial t}\hat{c}\down[\vec{k}]^{\dagger}  &=
[H,\hat{c}\down[\vec{k}]^{\dagger}].
\end{split}
\end{equation}

As it stands, the Hamiltonian in Eq. \eqref{eq:k-space hamiltonian} also contains pairs
with net momentum. According to the BCS approximation, such terms can
be dropped. This is formally achieved by setting \(\vec{k}' = 0\) in Eq. \eqref{eq:k-space
  hamiltonian}. With this simplification, the commutators are
\begin{equation}
\begin{split}
[H,\hat{c}\up[-\vec{k}]] &= -(\epsilon_{\vec{k}} -\mu)\hat{c}\up[-\vec{k}]
-\frac{U}{M}\hat{c}\down[\vec{k}]^\dagger\sum_{\vec{q}}\hat{c}\down[-\vec{k}-\vec{q}]\hat{c}\up[\vec{k}+\vec{q}]\\ 
[H,\hat{c}\down[\vec{k}]^\dagger] &= (\epsilon_{\vec{k}}
-\mu)\hat{c}\down[\vec{k}]^\dagger -\frac{U}{M}
\sum_{\vec{q}}\left(\hat{c}\up[\vec{k}+\vec{q}]^\dagger
\hat{c}\down[-\vec{k}-\vec{q}]^\dagger\right) \hat{c}\up[-\vec{k}]. 
\end{split}
\end{equation}
These equations are then linearized by replacing the sum
\(\sum_{\vec{q}}\hat{c}\down[-\vec{k}-\vec{q}]\hat{c}\up[\vec{k}+\vec{q}]\) with
its expectation value. When the chemical potentials for different
components are the same, this value is always independent of position in equilibrium.
\begin{equation}
\begin{split}
  \label{eq:linearization}
  -\frac{U}{M}\sum_{\vec{q}}\left\langle
    \hat{c}\down[-\vec{k}-\vec{q}]\hat{c}\up[\vec{k}+\vec{q}]\right\rangle &= \Delta\\
  -\frac{U}{M}\sum_{\vec{q}}\left\langle
    \hat{c}\up[\vec{k}+\vec{q}]^\dagger
\hat{c}\down[-\vec{k}-\vec{q}]^\dagger   \right\rangle &= \Delta^*.
\end{split}
\end{equation}
Because the overall phase does not play any role in the mean-field
theory, it is possible to select the order parameter \(\Delta\) to be
real. The equations of motion are thus reduced to
\begin{equation}
  -i\hbar\frac{\partial}{\partial
    t}\begin{pmatrix}\hat{c}\up[-\vec{k}]\\\hat{c}\down[\vec{k}]^\dagger
  \end{pmatrix} = \begin{pmatrix}-(\epsilon_{\vec{k}} - \mu)& \Delta
    \\\Delta & \epsilon_{\vec{k}}-\mu \end{pmatrix}\begin{pmatrix}\hat{c}\up[-\vec{k}]\\\hat{c}\down[\vec{k}]^\dagger
  \end{pmatrix}.
\end{equation}
These equations are decoupled with the standard
Bogoliubov-transformation, i.e. with a transformation that
diagonalizes the Hamiltonian
\begin{equation}
  \begin{pmatrix}\hat{c}\up[-\vec{k}]\\\hat{c}\down[\vec{k}]^\dagger\end{pmatrix}
  = \begin{pmatrix}u_{-\vec{k}} & v_{-\vec{k}}\\-v_{\vec{k}} &
    u_{\vec{k}} \end{pmatrix}\begin{pmatrix}\hat{\gamma}\up[-\vec{k}]\\\hat{\gamma}\down[\vec{k}]^\dagger \end{pmatrix}.
\end{equation}
For this transformation to be canonical, the new operators 
must obey the fermionic anticommutation rules. This leads to
the requirement that \(u_{\vec{k}}^2 + v_{\vec{k}}^2 = 1\). The
transformation that decouples the equations is given by 
\begin{equation}
\label{eq:uandv}
\begin{split}
  u_{\vec{k}}^2 &= \frac{1}{2}\left(1 + \frac{\epsilon_{\vec{k}} -
      \mu}{E_{\vec{k}}}\right) \\
  v_{\vec{k}}^2 &= \frac{1}{2}\left(1 - \frac{\epsilon_{\vec{k}} -
      \mu}{E_{\vec{k}}}\right) \\
   u_{\vec{k}}v_{\vec{k}} &= \frac{\Delta}{2E_{\vec{k}}},
\end{split}
\end{equation}
where \(E_{\vec{k}}\) is the quasiparticle dispersion, i.e.
\begin{equation}
  \label{eq:quasiparticle}
  E_{\vec{k}} = \sqrt{(\epsilon_{\vec{k}} -\mu)^2 + \Delta^2}.
\end{equation}
Inserting the transformed operators back to
Eq. \eqref{eq:linearization} gives the so called gap equation:
\begin{equation}
  \label{eq:gap}
    \Delta = -\frac{U}{M}\sum_{\vec{k}}\left\langle \hat{c}\down[-\vec{k}]\hat{c}\up[\vec{k}]\right\rangle=
    -\frac{U}{M}\sum_{\vec{k}}\bigg\langle\left(-v_{-\vec{k}}\hat{\gamma}\up[\vec{k}]^\dagger + u_{-\vec{k}}\hat{\gamma}\down[-\vec{k}] \right) \cdot\left(u_{\vec{k}}\hat{\gamma}\up[\vec{k}] + v_{\vec{k}}\hat{\gamma}\down[-\vec{k}]^\dagger \right)\bigg\rangle. 
\end{equation}
Because the quasiparticle operators \(\hat{\gamma}\) describe fermions, they
obey Fermi statistics. Therefore, in thermal equilibrium,
\begin{equation}
\label{eq:thermal}
  \begin{split}
    \langle \hat{\gamma}\up[\vec{k}]^\dagger \hat{\gamma}\up[\vec{k}]\rangle  &=
    f(E_{\vec{k}})\\  \langle\hat{\gamma}\down[-\vec{k}]
    \hat{\gamma}\down[-\vec{k}]^\dagger\rangle &= 1 - f(E_{-\vec{k}})\\
    \langle \hat{\gamma}\up[\vec{k}]^\dagger \hat{\gamma}\down[-\vec{k}]^\dagger
    \rangle &= 0\\
    \langle \hat{\gamma}\down[-\vec{k}]\hat{\gamma}\up[\vec{k}] \rangle &= 0,
  \end{split}
\end{equation}
and
\begin{equation}
  \label{eq:gap_final}
  \Delta = -\frac{U}{M}\sum_{\vec{k}} \frac{\Delta}{2E_{\vec{k}}}(1-2f(E_{\vec{k}})),
\end{equation}
where \(f\) is the Fermi distribution. On the other hand, the number of particles is equal to
\begin{equation}
  \label{eq:number}
  N = \sum_{\vec{k}} \left\langle \hat{c}\down[\vec{k}]^\dagger
  \hat{c}\down[\vec{k}] + \hat{c}\up[\vec{k}]^\dagger \hat{c}\up[\vec{k}]\right\rangle
\end{equation}
and using Eq. \eqref{eq:uandv} and Eq. \eqref{eq:thermal} this becomes the number equation,
\begin{equation}
N = 2\sum_{\vec{k}}\left(u_{\vec{k}}^2 f(E_{\vec{k}}) + v_{\vec{k}}^2 (1-f(E_{\vec{k}})) \right).
\end{equation}

Instead of the total number of particles, it is useful to deal with
the filling fraction, i.e. the total number of particles divided by the
number of lattice sites, \(N/M\). In our notation this ratio can have
values between \(0\) and \(2\). The former corresponds to no
particles, whereas the latter describes a full lattice with two atoms
of opposite spins at each site.


\section{Density response}
\label{sec:DR}
Density response describes how the total density of the system changes as a
result of a (sharp) change in the external potential. In the
linear response regime, in a homogenous systema and under equilibrium
conditions, it is possible to write
\begin{equation}
\delta\rho(\vec{k},\omega) = \chi(\vec{k},\omega)\delta V(\vec{k},\omega),
\end{equation}
where \(\delta V(\vec{k},\omega)\) is the change in the external potential
and \(\chi(\vec{k},\omega)\) is the density response. 

To calculate the response function \(\chi(\vec{k},\omega)\), we use the method of
the generalized random phase approximation 
(GRPA)~\cite{Anderson1958a,Cote1993a,Belkhir1994a}, following
the notation used by C{\^o}t{\'e} and Griffin~\cite{Cote1993a}.
Let us briefly summarize the final results of their
derivations. First, define two matrices, \(\widetilde{A}\) and \(\widetilde{B}\):

\begin{equation}
\widetilde{A}(\vec{k}) = \begin{pmatrix} u_{\vec{k}}^2 &
  u_{\vec{k}}v_{\vec{k}}\\-u_{\vec{k}}v_{\vec{k}} & -v_{\vec{k}}^2  \end{pmatrix}
\end{equation}

\begin{equation}
\widetilde{B}(\vec{k}) = \begin{pmatrix} v_{\vec{k}}^2 &
  -u_{\vec{k}}v_{\vec{k}}\\u_{\vec{k}}v_{\vec{k}} & -u_{\vec{k}}^2  \end{pmatrix}.
\end{equation}

Then define a \(4\times4\)-matrix, \(L^0\):

\begin{equation}
\label{L-matriisi}
L^0 = \begin{pmatrix}L_{1111} & L_{1121} & L_{1211} &
  L_{1221}\\L_{1112} & L_{1122} & L_{1212} & L_{1222}\\
  L_{2111} & L_{2121} & L_{2211} & L_{2221}\\
  L_{2112} & L_{2122} & L_{2212} & L_{2222}
 \end{pmatrix},
\end{equation}
with matrix elements given by 
\begin{equation}
\label{eq: L-matriisi}
\begin{split}
L_{ijkl}(\vec{q},\omega) & = 
\sum_{\vec{q}'}\bigg(\frac{\widetilde{A}_{ij}(\vec{q}+\vec{q}')\widetilde{A}_{kl}(\vec{q}')
}{E_{\vec{q}'}-E_{\vec{q}+\vec{q}'}+\hbar(\omega+i\delta)}\left(f - f' \right)
+ \frac{\widetilde{B}_{ij}(\vec{q}+\vec{q}')\widetilde{B}_{kl}(\vec{q}')
}{E_{\vec{q}'}-E_{\vec{q}+\vec{q}'}-\hbar(\omega + i \delta)}\left(f - f' \right)\\
+& \frac{\widetilde{A}_{ij}(\vec{q}+\vec{q}')\widetilde{B}_{kl}(\vec{q}')
}{E_{\vec{q}'}+E_{\vec{q}+\vec{q}'}-\hbar(\omega + i \delta)}\left(f + f' -1
\right)
+ \frac{\widetilde{B}_{ij}(\vec{q}+\vec{q}')\widetilde{A}_{kl}(\vec{q}')
}{E_{\vec{q}'}+E_{\vec{q}+\vec{q}'}+\hbar(\omega + i \delta)}\left(f + f' -1 \right)
\bigg),
\end{split}
\end{equation}
where \(f\) is the Fermi function at \(E_{\vec{q}'}\), \(f'\) is
the Fermi function at \(E_{\vec{q}'-\vec{q}}\), and the convergence
factor \(\delta\) is put to \(0\) after the calculation.

Due to symmetries, only six of the 16 elements in \(L^0\) are
actually independent~\cite{Cote1993a}. These elements can be denoted as

\begin{equation}
\begin{split}
a &= L_{1111} = L_{2222}\\
b &= L_{1212} = -L_{1221} = -L_{2112} = L_{2121}\\
c &= L_{1112} = -L_{1121} = L_{1222} = -L_{2122}\\
\bar{c} &= L_{1211} = -L_{2111} = L_{2212} = -L_{2221}\\
d &= L_{1122}\\
\bar{d} &= L_{2211}.
\end{split}
\end{equation}
The number of independent components can be reduced further
by assuming weak coupling, as is done in \cite{Cote1993a}. In this limit,
\(\bar{c} = -c\) and \(\bar{d} = d\). However, in order to probe also more
strongly interacting systems, we do
not make this assumption. We then define a \(4\times1\) vector \(\hat{L}\),
\begin{equation}
\hat{L} = \begin{pmatrix} a - b \\ 2c \\ 2\bar{c} \\ a - b\end{pmatrix},
\end{equation}
and solve the linear algebra problem 
\begin{equation}
\left(\mathbbm{1} + UL^0 \right)\vec{x} = \hat{L},
\end{equation}
where \(U\) is the interaction strength in the Hamiltonian. Finally,
from the solution \(\vec{x}\)
we deduce the density response function from \(\vec{x}\): 

\begin{equation}
\chi(\vec{k},\omega) = \frac{x_1 + x_4}{1 - U(x_1 + x_4)}.
\end{equation}

We consider the dynamic structure factor instead of the
explicit density response, because the former is 
measurable using, for example,
Bragg spectroscopy~\cite{Stenger1999a}. The dynamic structure factor
is given by
\begin{equation}
  \label{eq:S}
  S(\vec{k},\omega) = -\frac{1}{\pi}\text{Im}\chi(\vec{k},\omega).
\end{equation}

Formally, the convergence factor \(\delta\) defined in Eq. \eqref{eq:
  L-matriisi} is put to \(0\) after the
calculations. However, for the sake of illustration, we use a small
finite value for \(\delta\) in the figures. This gives rise to a finite linewidth in the
density response.

\begin{figure}
\includegraphics{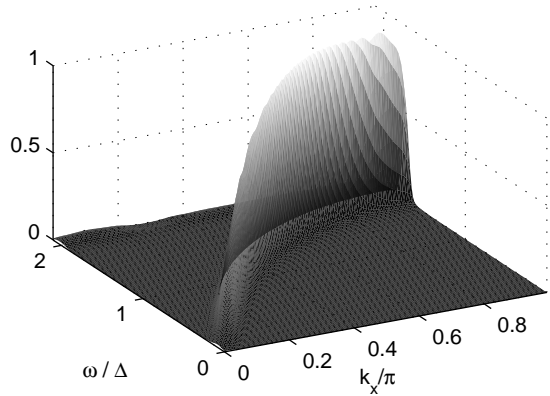}
\caption{The Anderson-Bogoliubov phonon in an cubic lattice. The
  \(x\)-axis is the \(x\)-component of the momentum vector in the
  units of inverse lattice spacing, the \(y\)-axis is energy in the
  units of the pairing gap divided by \(\hbar\), and the \(z\)-axis shows the dynamical
  structure factor from Eq. \eqref{eq:S}, in arbitrary units.}
\label{fig:A-B}
\end{figure}

Because the experimentally relevant lattice sizes are relatively
small, only in the order of 100 sites per dimension, it is possible to
explicitly calculate, i.e. without approximating by integrals, the
sums over all lattice sites in the BCS-theory and the density response. 
For concreteness all our results are calculated for 
$^6$Li atoms at zero temperature 
and with a laser wavelength \(\lambda = 1030\)~nm.
Restriction into $T=0$ is accurate
when the energy gap is much larger than the temperature.
Our
numerical calculation correctly reproduces the Anderson-Bogoliubov 
phonon, see Fig.~\ref{fig:A-B}. This mode is gapless and is expected
on general grounds when the continuous U(1) symmetry is 
broken \cite{Anderson1958a}.

\begin{figure}
\includegraphics{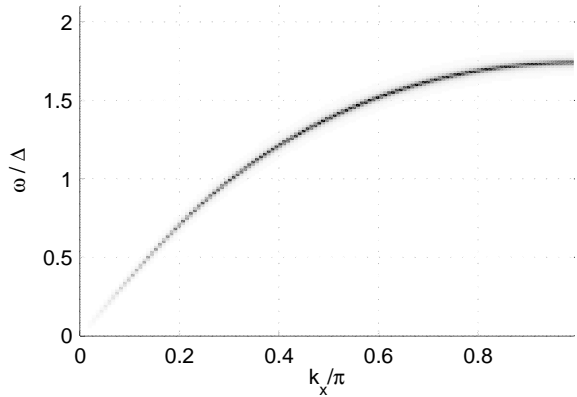}
\caption{The dispersion of the Anderson-Bogoliubov phonon. The \(x\)-axis is the \(x\)-component of the momentum vector in the
  units of inverse lattice spacing and the the \(y\)-axis is energy in the
  units of the pairing gap divided by \(\hbar\).}
\label{fig:A-B dispersion}
\end{figure}

\section{Speed of sound in a cubic 3D lattice}
\label{sec:SoS}
The density response in \((\vec{k},\omega)\) space
(see Fig. \ref{fig:A-B dispersion}) gives the dispersion of sound in
the lattice. The long wavelength limit of the dispersion is linear and
therefore the speed of sound is independent of momentum in that
limit. 
For higher momenta the response saturates to \(2\Delta\). 
In general the speed of sound in a weakly interacting Fermi gas is (to
the leading order) \(v_F/\sqrt{3}\), where \(v_F\) is the Fermi velocity of the
system. However, since the Fermi surface in a lattice is not a sphere,
the Fermi velocity is not a uniquely defined number, but depends on
the direction. With the definition
\begin{equation}
  \label{eq:fermivelocity}
  v_{Fx} =
  \frac{1}{\hbar}\left.\frac{d\epsilon_{\vec{k}}}{dk_x}\right|_{\vec{k}=k_{Fx}} = \frac{2J_x d}{\hbar}\sqrt{\left|\frac{E_F}{J_x}-\frac{1}{4}\left(\frac{E_F}{J_x}\right)^2\right|},
\end{equation}
our results agree with the free space result in the low density limit.

\begin{figure}
\includegraphics{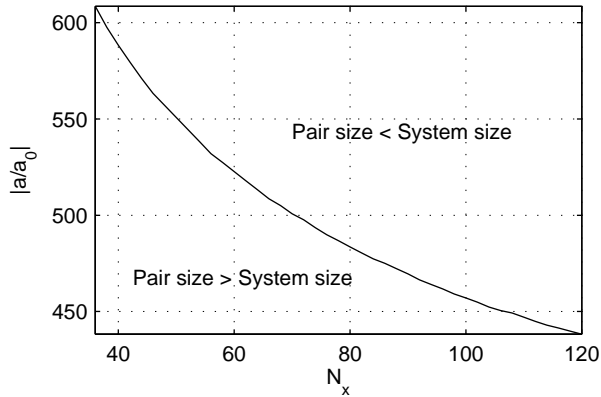}
\caption{Pair size compared with the system size. The \(x\)-axis is
  the number of the lattice sites in the \(x\)-direction and the
  \(y\)-axis is the absolute value of the scattering length, in Bohr
  radii. For the density response
to be experimentally observable, the coherence length, i.e. the size
of a Cooper pair, must be smaller than the system. The
linear sound dispersion is clearly visible above the solid line.}
\label{fig:pair size}
\end{figure}

In order to experimentally observe the density response in a finite
system one must overcome some problems. In particular, the region of
\(\vec{k}\)-space where the dispersion relation is linear scales with
the inverse of the coherence length (or Cooper pair size) i.e. \(\delta
k \sim 1/\xi\). In the weakly interacting system, Cooper pairs become
very large and the density response saturates very quickly to
\(2\Delta\). In the lattice the smallest non-zero wavevector has the
magnitude of \(k_x^{\text{min}} = 2\pi/N_x\lambda\), where \(N_x\) is the lattice size in
the \(x\)-direction. This should be much smaller than the linear
region of the dispersion. In other words the system size must be much
larger than the coherence length. This leads to a
condition for the minimum system size where the observation of density
response is possible, which is shown in Fig.~\ref{fig:pair size}.

\subsection{Speed of sound and the van Hove singularity}
\begin{figure}
\includegraphics{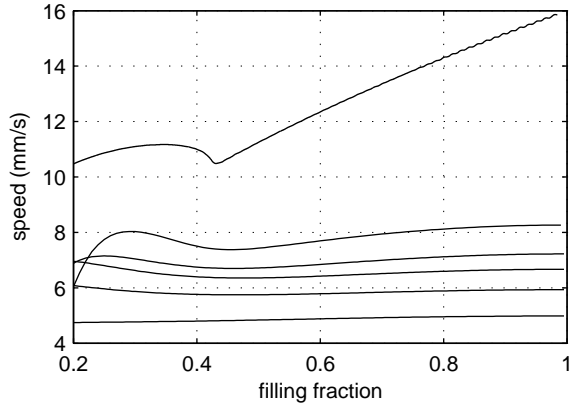}
\caption{The speed of sound as a function of the lattice filling
  fraction, for the non-interacting Fermi gas (top curve) and for
  interacting Fermi gases with different scattering lenghts: (in Bohr
  radii, from top to bottom) \(-750\), \(-1000\), \(-1200\), \(-1500\), and
  \(-2000\). The speed has a local minimum at a point where \(E_F =
  4J\), but this minimum is smoothened out as interactions become
  stronger.}
\label{fig:isotropic speed by filling}
\end{figure}

Unlike in free space, in a lattice the speed of sound changes with the
density of the atoms in a non-trivial way. Fig.~\ref{fig:isotropic speed by filling} shows the speed as
a function of the filling fraction with different interaction
strengths. The sound velocity is typically on the order of ${\rm mm/s}$
and is thus experimentally measurable.
With small scattering lengths and in a cubic lattice, the speed of sound as a function of the
filling fraction is
non-monotonous and exhibits a
local minimum around the point where \(E_F = 4J\). For non-interacting
fermions this minimum is a sharp cusp corresponding to the van
Hove singularity in the density of states. The cusp appears when the
Fermi surface in the direction of the sound reaches the divergence in the
density of states. The cusp is visible in Fig. \ref{fig:isotropic speed by filling} where we show
the isothermal sound velocity
\begin{equation}
c_T=\sqrt{\frac{\partial P}{\partial \rho}},
\end{equation}
where \(P\) is the pressure and \(\rho\) is the mass density of the
gas, calculated for the ideal Fermi gas in a finite sized lattice. 

However, interactions smear the Fermi surface by
broadening its edge by an amount of the pairing gap, \(\Delta\). This
smoothens out the cusp caused by the singularity. In fact, as is clear
from the Fig. \ref{fig:isotropic speed by filling}, the effect will
vanish completely with strong enough interactions.

The finite size effects introduce some modulations, but it is clear
that a non-zero gap shifts the
minimum of the sound velocity into higher filling fractions.
The non-zero gap also lowers the sound velocity and this effect
becomes more pronounced at higher filling fractions.
For a weakly interacting system with a very low filling fraction and
an infinite lattice, the speed of sound is given by the
analytical result~\cite{Anderson1958a}
\begin{equation}
c=\frac{v_F}{\sqrt{3}}\sqrt{1+\frac{U(6\pi^2n)^{1/3}}{4\pi^2J_xd^2}}.
\end{equation}

\begin{figure}
\includegraphics{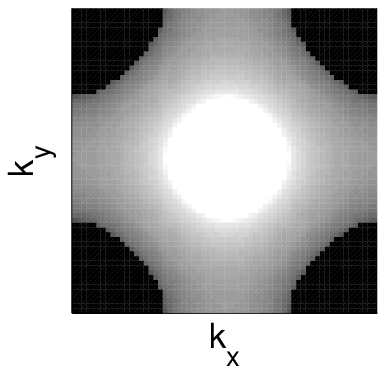}
\includegraphics{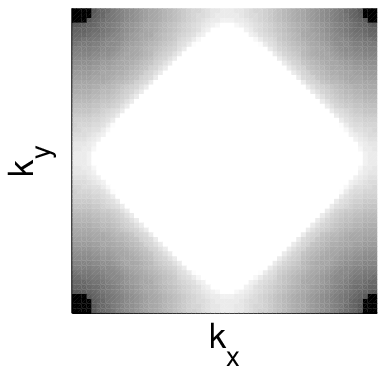}
\includegraphics{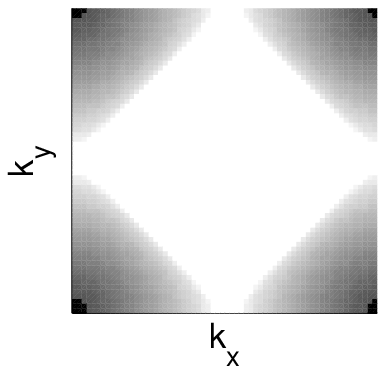}
\includegraphics{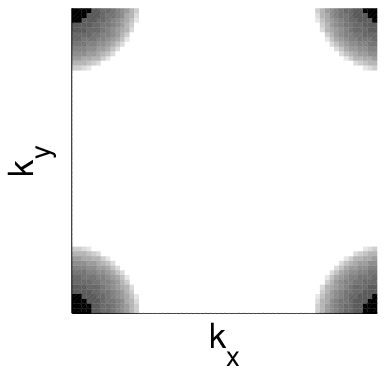}
\caption{The Fermi surface of the lattice with different densities, as seen from
the z-direction. The axis are the \(x\)- and \(y\)-components of the
momentum vector. The surfaces above were plotted with filling fractions
$0.2$, $0.44$, $0.47$, and $0.8$, from left to right. The colour coding:
white corresponds to the Fermi surface reaching the edge of the Brillouin
zone in the $z$-direction, whereas the black area is outside the surface.
Note that the presence of the excitation gap spreads the Fermi surface,
wheras for a non-interacting system, the edges of the Brillouin zone are
not reached until the filling fraction exceeds 0.42.}
\label{fig:fermipinnat}
\end{figure}

The location of the van Hove singularity is also the point where the Fermi surface
of the system becomes disconnected in a non-interacting system. In the
interacting case, the Fermi surface is smoothened, see Fig. \ref{fig:fermipinnat}.

\subsection{Speed of sound as a function of interaction strength}
\begin{figure}
\includegraphics{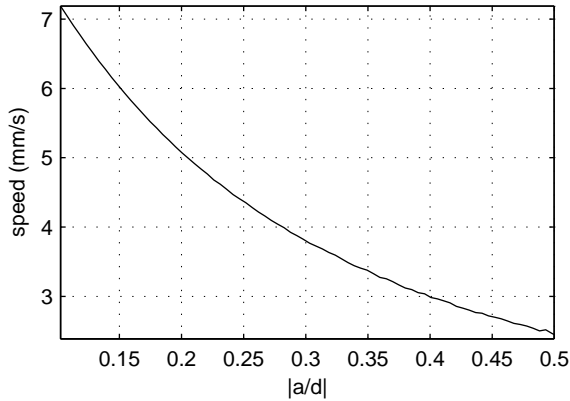}
\caption{The speed of sound as a function of the scattering
  length \(a\). Here \(d\) is the lattice spacing, the
  filling fraction is \(0.5\), and the lattice is cubic, with
  \(s_x,s_y,s_z = 2.5\).}
\label{fig:speed by scattering}
\end{figure}

Increasing the interactions between atoms (i.e. increasing the
scattering length) tends to reduce the sound velocity, as can be seen in
Fig. \ref{fig:speed by scattering}. This behaviour is expected on the
grounds that interaction strength increases the density of states in general,
which increases the compressibility of the system, which in turn reduces
the speed of sound. Note that in contrast, for a Bose-Einstein condensate with a
positive scattering length the sound velocity actually
increases as interactions become stronger.

\section{Non-cubic lattices and dimensional crossover}
\label{sec:AIS}
We now study density response and the speed of sound in
non-cubic lattices. By varying the lattice heights \(s_x\), \(s_y\)
and \(s_z\), i.e. by changing the different laser
intensities, the cubic symmetry of the lattice can be broken. For concreteness, we
choose \(s_x = s_y\). With this simplification, there are basically
two distinct cases, namely \(s_x > s_z\) and \(s_x < s_z\). In the
former case the atoms are more free to move in the \(z\)-direction than in the \(xy\)-plane. This means that the lattice resembles a
square lattice arrangement of coupled one-dimensional tubes. The
latter corresponds to a situation where movement is more restricted in
the \(z\)-direction. In this case the lattice can thought to be a pile
of connected two-dimensional planes (``pancakes'').

\begin{figure}
\includegraphics{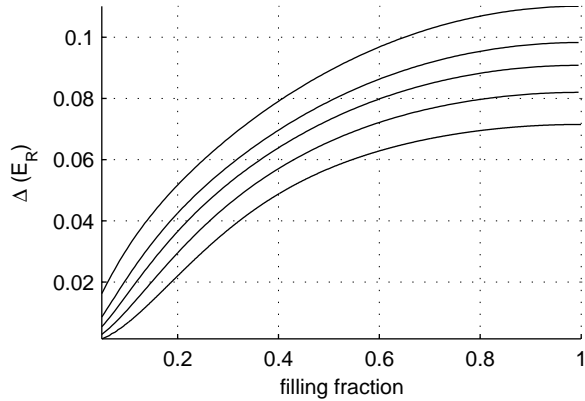}
\caption{The energy gap as a function of the filling fraction with \(s_x = s_y < s_z\),
  corresponding to quasi-two-dimensional pancakes along the
  \(xy\)-plane. The different curves are for different ratios of
  \(s_x/s_z\): (from top to bottom) \(0.5\), \(0.62\), \(0.71\),
  \(0.83\), and \(1.0\) (cubic case). There is no qualitative
  difference in the different curves. Here the scattering
  length \(a\) is all the time \(-1000 a_0\). }
\label{fig:gaps}
\end{figure}

In either case, the energy gap \(\Delta\), does not show any
qualitative change when the symmetry decreases: \(\Delta\) merely grows
as either \(s_x\) (and \(s_y\)) or \(s_z\) is increased, since tighter
confinement effectively increases interaction, see Fig.~\ref{fig:gaps}. However,
there is a clear signal of a dimensional crossover in the speed of sound. In the two-dimensional limit, i.e. when the tunneling in the
\(z\)-direction is suppressed (pancakes), the local minimum of the speed of sound
vanishes, as can be seen from Fig. \ref{fig:pancakes speed by
  filling}. Also it can be seen that the speed of sound is significantly
larger in the direction parallel to the planes than orthogonal to
them. 

\begin{figure}
\includegraphics{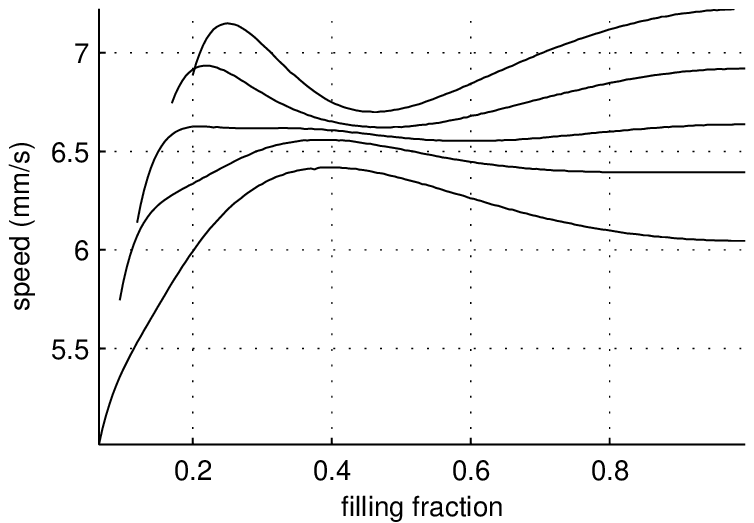}
\includegraphics{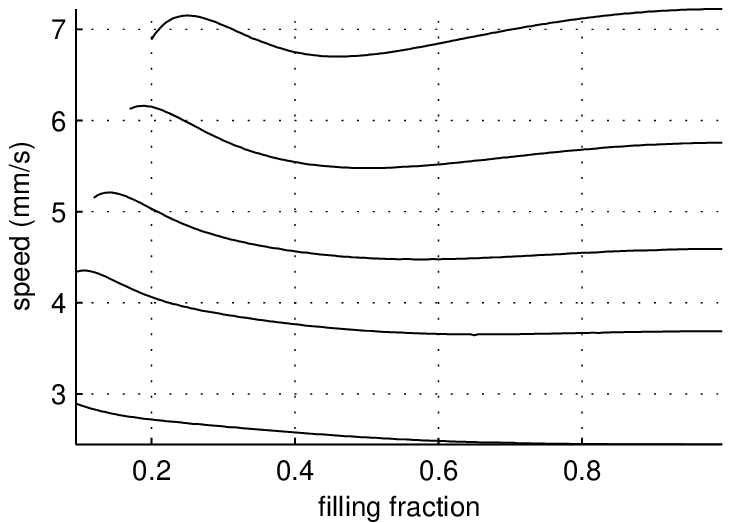}
\caption{The speed of sound as a function of the lattice filling
  fraction in non-cubic lattices with \(s_x = s_y < s_z\),
  corresponding to quasi-two-dimensional pancakes along the \(xy\)-plane. The left figure shows
  the speed in the plane, i.e. parallel to the pancakes, and
  the one on the right the speed orthogonal to the them, i.e. along the \(z\)-axis. The different curves
  are for different ratios of \(s_x/s_z\): (from top to bottom) \(1.0\) (cubic case),
  \(0.83\), \(0.71\), \(0.62\), and \(0.5\). Here the scattering
  length \(a\) is all the time \(-1000 a_0\).}
\label{fig:pancakes speed by filling}
\end{figure}

In the opposite case of stronger tunneling in the \(z\)-direction than in the
\(xy\)-plane (1D tubes), the local minimum in the speed of sound also
disappears. However, as is apparent from Fig. \ref{fig:tubes speed by
  filling}, in this one-dimensional limit the speed of sound
is again qualitatively different from both the one in the three-dimensional cubic
system and the one in the non-cubic system in the two-dimensional limit.

\begin{figure}
\includegraphics{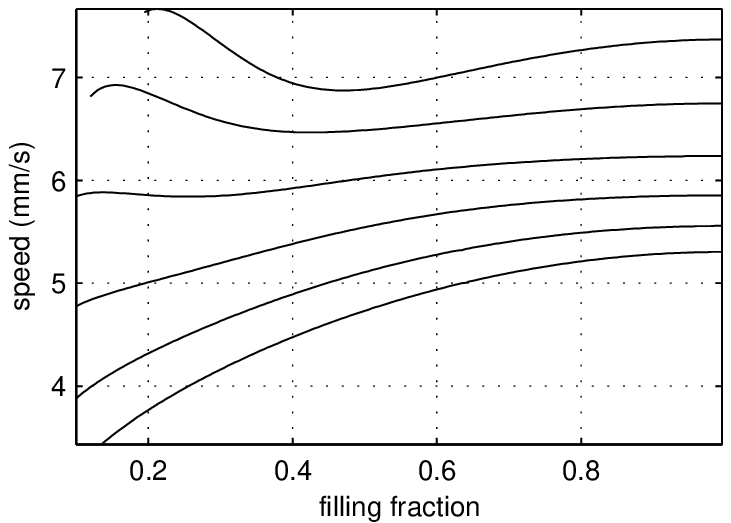}
\includegraphics{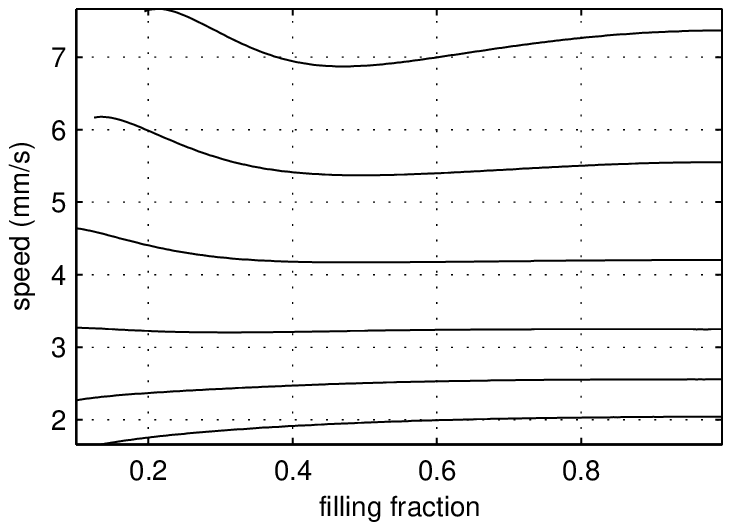}
\caption{The speed of sound as a function of the lattice filling
  fraction in non-cubic lattices with \(s_x = s_y > s_z\),
  corresponding to quasi-one-dimensional tubes along the
  \(z\)-axis. The figure on the left shows
  the speed in the \(z\)-direction, i.e. parallel to the tubes, and
  the right one the speed orthogonal to the them, i.e. in the \(xy\)-plane. The different curves
  are for different ratios of \(s_x/s_z\): (from top to bottom) \(1.0\) (cubic case),
  \(1.2\), \(1.4\), \(1.6\), \(1.8\), and \(2.0\). Here the background scattering
  length \(a\) is all the time \(-1000 a_0\).}
\label{fig:tubes speed by filling}
\end{figure}

\section{Conclusions}
\label{sec:conclusions}

In this paper, we have studied the velocity
of sound in cubic and non-cubic three-dimensional optical
lattices. We have investigated how the van Hove
singularity of the free Fermi gas is smoothened by interactions and
eventually vanishes when interactions are strong enough. In
non-cubic lattices we have shown that although the energy gap has a
simple behaviour as a function of symmetry, the speed of sound shows
qualitatively different behaviour over the crossover, and provides a
clear experimentally observable signature of a dimensional crossover
both in the 1D and 2D limits. 

\begin{acknowledgments}
This work was supported by the Finnish Cultural Foundation, the
Academy of Finland (project numbers 106299, 7205470), EUROHORCs (EURYI award), and the QUPRODIS
project of EU. 
\end{acknowledgments}


\end{document}